\documentclass[useAMS, referee]{mn2e}
\usepackage{graphicx}
\usepackage{amsmath}

\title[The X-ray Pulsar 4U 1907+09]{A Comprehensive Study on \emph{RXTE} \& \emph{INTEGRAL} Observations of the X-ray Pulsar 
4U 1907+09}
\author[\c{S}. \c{S}ahiner, S. \c{C}. \.{I}nam and A. Baykal ]
{\c{S}. \c{S}ahiner$^{1}$\thanks{E-mail: seyda@astroa.physics.metu.edu.tr (\c{S}\c{S}); inam@baskent.edu.tr (S\c{C}\.{I}); 
altan@astroa.physics.metu.edu.tr (AB)}, S. \c{C}. \.{I}nam$^{2}$\footnotemark[1] and A. Baykal$^{1}$\footnotemark[1] \\
$^{1}$Physics Department, Middle East Technical University, 06531 Ankara, Turkey\\
$^{2}$Department of Electrical and Electronics Engineering, Ba\c{s}kent University, 06530 Ankara, Turkey}

\begin{document}

\date{Received 2011}

\pagerange{\pageref{firstpage}--\pageref{lastpage}} \pubyear{2011}

\maketitle

\label{firstpage}

\begin{abstract}

We analyse \emph{INTEGRAL} (between 2005 October and 2007 November) and \emph{RXTE} (between 2007 June and 2011 March) 
observations of the accretion powered pulsar 4U 1907+09. From \emph{INTEGRAL} IBIS-ISGRI and \emph{RXTE}-PCA observations, 
we update pulse period history of the source. We construct power spectrum density of pulse frequencies and find that 
fluctuations in the pulse frequency derivatives are consistent with the random walk model with a noise strength of 
$1.27\times10^{-21}$ Hz s$^{-2}$. From the X-ray spectral analysis of \emph{RXTE}-PCA observations, we find that Hydrogen 
column density is variable over the binary orbit, tending to increase just after the periastron passage. We also show that 
the X-ray spectrum gets hardened with decreasing X-ray flux. We discuss pulse-to-pulse variability of the source near dipping 
ingress and egress. We find that the source more likely undergoes in dipping states after apastron until periastron when 
the accretion from clumpy wind might dominate so that occasional transitions to temporary propeller state might occur.
\end{abstract}

\begin{keywords}
accretion, accretion discs -- stars: neutron -- pulsars: individual: 4U 1907+09 -- X-rays: binaries.
\end{keywords}

\section{Introduction}

The X-ray source 4U 1907+09 was discovered in early 1970s by the \emph{Uhuru} survey (Giaconni et al. 1971). The system is 
an High mass X-ray binary (HMXB) that contains an X-ray pulsar accreting material from its blue supergiant companion star. 
The pulsar has an eccentric (\emph{e} $\sim$ 0.28) orbit around its companion and the orbital period of the binary system is 
$\sim$8.3753 days (In't Zand et al. 1998). 

The orbital profile of 4U 1907+09 exhibits two flares per orbit separated by $\simeq0.5$ orbital phase; a large primary at 
the periastron and the small secondary at the apastron (Marshall \& Ricketts 1980, In't Zand et al. 1998). The presence of the two 
phase-locked flares had led to the suggestion that the compact object passes through a circumstellar disk of matter around 
the equatorial plane of a B\emph{e} type stellar companion (Iye 1986, Cook \& Page 1987). However the position of 4U 1907+09 
in the Corbet diagram (Corbet 1984) indicates that the companion is probably an OB type supergiant. Latest optical 
(Cox et al. 2005) and infrared (Nespoli et al. 2008) observations actually showed that the companion could be classifed as 
an O8 - O9 Ia supergiant with a mass loss rate of $\dot{M} = 7 \times 10^{-6}$ $M_\odot$ yr$^{-1}$, and a lower limit of 
distance $\sim$5 kpc. 

The spin period of 4U 1907+09 was first measured as $\sim$437.5 s using \emph{Tenma} observations (Makishima et al. 1984). 
The pulse profile is known to be double peaked with variable amplitude. The profile is insensitive to the energies below 20 keV, 
whereas dramatic changes are present above 20 keV (In't Zand et al. 1998). Historical period measurements confirm that the 
pulsar had been steadily spinning down since its discovery until 1998 with an average rate of 
$\dot{\nu} = -3.54 \times 10^{-14}$ Hz s$^{-1}$ (Cook \& Page 1987, In't Zand et al. 1998, Baykal et al. 2001, 
Mukerjee et al. 2001). Afterwards, \emph{RXTE} observations in 2001 showed that the spin rate was lowered by a factor of 
$\sim$0.60 (Baykal et al. 2006). From \emph{INTEGRAL} observations, it was reported that the spin period had reached to a 
maximum of $\sim$441.3 s then, a torque reversal occurred and the source began to spin up with a rate of 
$2.58 \times 10^{-14}$ Hz s$^{-1}$ after 2004 May (Fritz et al. 2006). Recent measurements with \emph{RXTE} presented in 
\.{I}nam et al. (2009a) and this paper have revealed that 4U 1907+09 has returned spin down trend with a rate of 
$-3.59 \times 10^{-14}$ Hz s$^{-1}$, which is close to the previous steady spin down rate. This implies that another torque 
reversal should have been taken place before 2007 June. 

4U 1907+09 was identified as a variable X-ray source showing irregular flaring and dipping activities, $\sim$20\% 
observations of which were reported to be in dip state with no detectable pulsed emission (In't Zand et al. 1997). The 
typical duration of the dips was found to vary between few minutes to 1.5 hours. Variations in the X-ray 
flux were accepted to be the evidences of instability in the mass accretion rate where the dipping states are associated 
with the cessation of the accretion from an inhomogeneous wind of the companion star. 

X-ray spectra of 4U 1907+09 were basically described by a power law with photon index $\sim$1.2 and an exponential cutoff at 
$\sim$13 keV (Schwartz et al. 1980, Marshall \& Ricketts 1980, Makishima et al. 1984, Cook \& Page 1987, Chitnis et al. 1993, 
Roberts et al. 2001, Coburn et al. 2002, Baykal et al. 2006, Fritz et al. 2006). The continuum was found to be modified by 
highly variable Hydrogen column density (n$_H$) over the binary orbit, between $1 \times 10^{22}$ cm$^{-2}$ and 
$9 \times 10^{22}$ cm$^{-2}$ as a consequence of inhomogeneous accretion via dense stellar wind (In't Zand et al. 1997). 
A narrow spectral line around 6.4 keV corresponding to \emph{Fe} K$\alpha$ emission produced by the fluorescence of matter 
surrounding the pulsar was also observed in the spectra. The detailed determination of \emph{Fe} K emission complex has been 
recently reported using \emph{Suzaku} observations, wherein \emph{Fe} K$\beta$ emission has been detected for the 
first time (Rivers et al. 2010). Observations with \emph{Ginga} (Mihara 1995, Makishima et al. 1999) and \emph{BeppoSAX} 
(Cusumano et al. 1998) exhibited cyclotron resonant scattering features (CRSFs) at higher energies. Line energies of the 
fundamental and the second harmonic cyclotron lines were found to be $\sim$19 keV and $\sim$39 keV respectively, implying a 
surface magnetic field strength of $2.1 \times 10^{12}$ G (Cusumano et al. 1998).

In this paper, we present timing and spectral analysis of \emph{RXTE} monitoring observations of 4U 1907+09 between 2007 
June and 2011 March. Selected \emph{INTEGRAL} pointing observations are analyzed to cover the gap in spin period history. 
We describe the observations in Section 2. Pulse timing analysis is reported in Section 3. In Section 4, spectral results 
are discussed. In Section 5, we focus on the dipping states of the source. Finally, we summarize our results in Section 6.

\section{Observations}

\subsection{\emph{RXTE}}

One of the main instruments on-board \emph{RXTE} is the \emph{Proportional Counter Array} (PCA) which consists of five 
co-aligned identical proportional counter units (PCUs) pointed to the same location in the sky (Jahoda et al. 1996). The 
field of view (FOV) of PCA at full width at half maximum (FWHM) is about 1$\degr$ and the effective area of each detector is 
approximately 1300 $cm^{2}$. PCA operates in the energy range 2 - 60 keV with an energy resolution of 18\% at 6 keV.

In this paper, we present timing and spectral analysis of 98 pointed \emph{RXTE}-PCA observations of 4U 1907+09 between
2007 June and 2011 March each with an exposure of $\sim$2 ks (see Table \ref{obsID}). Preliminary timing analysis were 
already performed and corresponding spin-rate measurements were presented before by \.{I}nam et al. (2009a) and \c{S}ahiner et 
al. (2011). In this paper, we extend this analysis by using observations with a longer time span and performing a more 
detailed timing analysis as well as presenting our spectral analysis results. 

Although number of active PCUs during the observations of 4U 1907+09 varies between one and three, data obtained from 
PCUs 0 and 1 are not appropriate for spectral analysis due to increased background levels as a result of the loss of their 
propane layers. Consequently, the only data products taken into consideration during spectral analysis belong to PCU 2 in 
order to avoid probable problems due to calibration differences between the detectors. The loss of propane layers of PCU 0 
and PCU 1 do not affect high resolution timing, therefore no PCU selection is done for the timing analysis.

\begin{table}
  \caption{Log of \emph{RXTE} observations of 4U 1907+09.}
  \label{obsID}
  \center{\begin{tabular}{lccl}
  \hline	
RXTE 		& Time 	& Number of 	& Exposure \\
proposal ID 	& (MJD) & observations 	& (ks) \\	
 \hline
 93036	& 54280 - 54825 & 39 & 74.1$^{a,b}$ \\
 94036	& 54839 - 55184 & 24 & 45.3$^b$ \\
 95350	& 55208 - 55555 & 26 & 49.4 \\
 96366	& 55571 - 55638 & 9  & 47.2 \\
\hline
\end{tabular}} \\
\begin{flushleft}
$^{a}$ Results from the first 30 observations were published before by \.{I}nam et al. (2009a).

$^{b}$ Preliminary timing analysis results were presented before by \c{S}ahiner et al. (2011).
\end{flushleft}
\end{table} 

The standard software tools of \verb"HEASOFT v.6.10" are used for the analysis of PCA data. Filter files are produced to 
apply constraints on the data such that; the times when elevation angle is less than 10$\degr$, offset from the source 
is greater than 0.02$\degr$ and electron contamination of PCU2 is greater than 0.1 are excluded. Data modes examined for 
the spectral and light curve extraction are 'Standard2f' and 'GoodXenon' modes respectively. Background spectra and light 
curves are generated by the latest PCA background estimator models supplied by the \emph{RXTE} Guest Observer Facility (GOF)
, \verb"Epoch 5C".

\subsection{\emph{INTEGRAL}}

The \emph{INTEGRAL} observations analysed in this paper are obtained from the \emph{INTEGRAL} Science Data Centre (ISDC) 
archive. All publicly available pointing observations subsequent to the previous study of 4U 1907+09 (Fritz et al. 2006) 
are selected considering the good IBIS-ISGRI times to be above 1 ks. Selected observations, in which the source is in the 
FOV with an off-set angle smaller than 5$^\circ$, are held on between 2005 October and 2007 November. The data consist of a total of 611 science windows (SCWs) 
(each $\sim$3 ks) within revolutions 366 and 623. 

The data products of IBIS-ISGRI detector on-board \emph{INTEGRAL} are reduced for the analysis. \emph{Imager on Board the 
INTEGRAL Satellite} (IBIS) is a coded mask instrument, which has a fully coded FOV of 8$\degr.3 \times 8\degr.0$ and 
12$\arcmin$ angular resolution (FWHM) (Ubertini et al. 2003). \emph{INTEGRAL Soft Gamma-Ray Imager} (ISGRI) is the upper 
layer of the IBIS instrument which operates in the energy range 15 keV - 1 MeV with an energy resolution of 8\% at 60 keV 
(Lebrun et al. 2003). The data reduction is performed by the software \verb"OSA v.7.0". The standard pipeline processing 
comprises gain correction, good-time handling, dead-time derivation, background correction and energy reconstruction. Images 
in two energy bands (20 - 40 keV and 40 - 60 keV) are produced from IBIS-ISGRI data with the use of an input catalogue 
consisting strong sources in the FOV: Ser X-1, XTE J1855-026, 4U 1909+07, SS 433, IGR J19140+0951, GRS 1915+105 and 4U 
1907+09. The background maps provided by the ISGRI team are used for background correction. Light curves are created by the 
tool \verb"II_LIGHT" which allows high resolution timing. 

\section{Timing Analysis}

\begin{figure}
 \center{\begin{tabular}{cc}
   \includegraphics[height=6.2cm, angle=270]{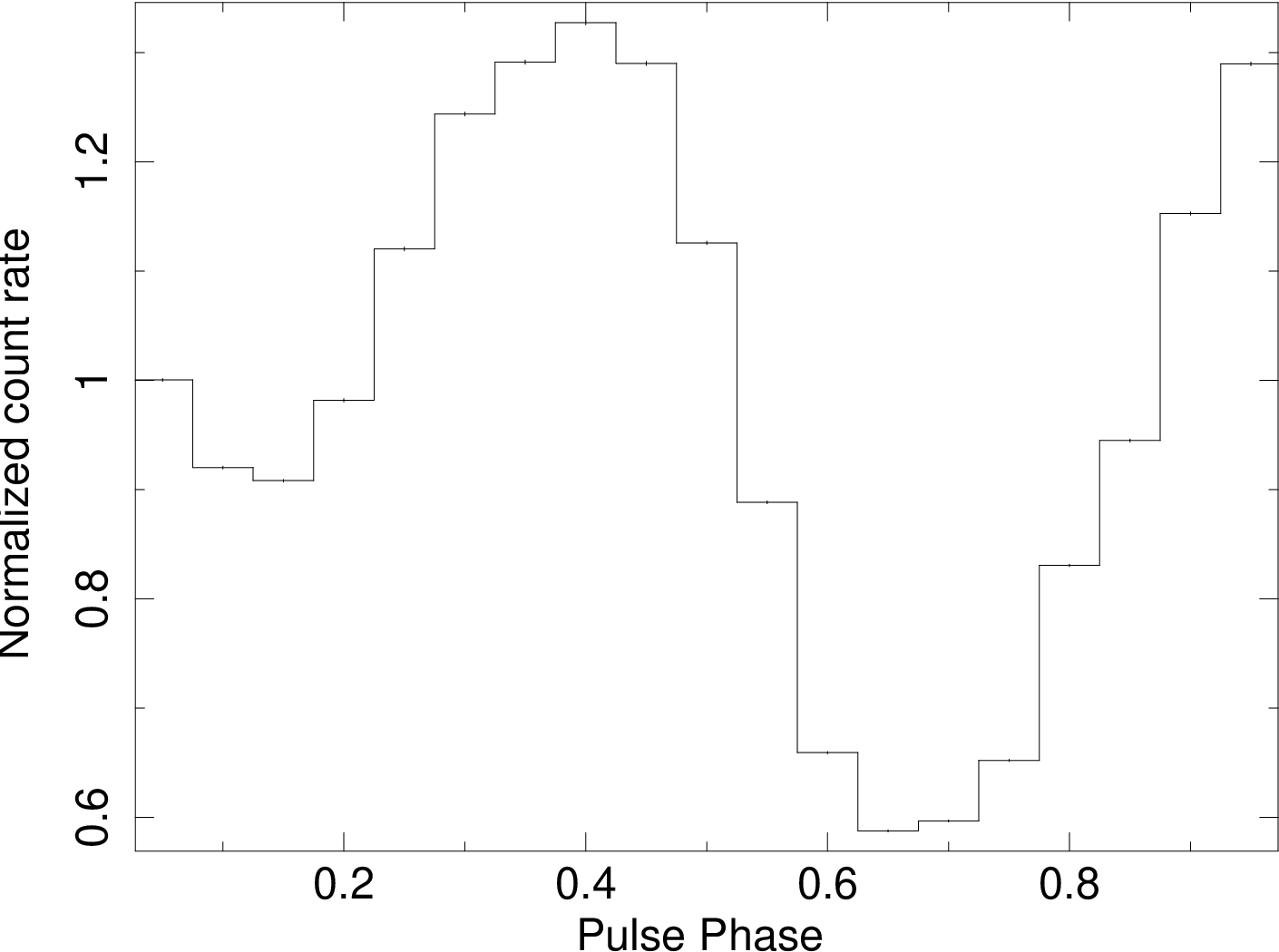} & \includegraphics[height=6.2cm, angle=270]{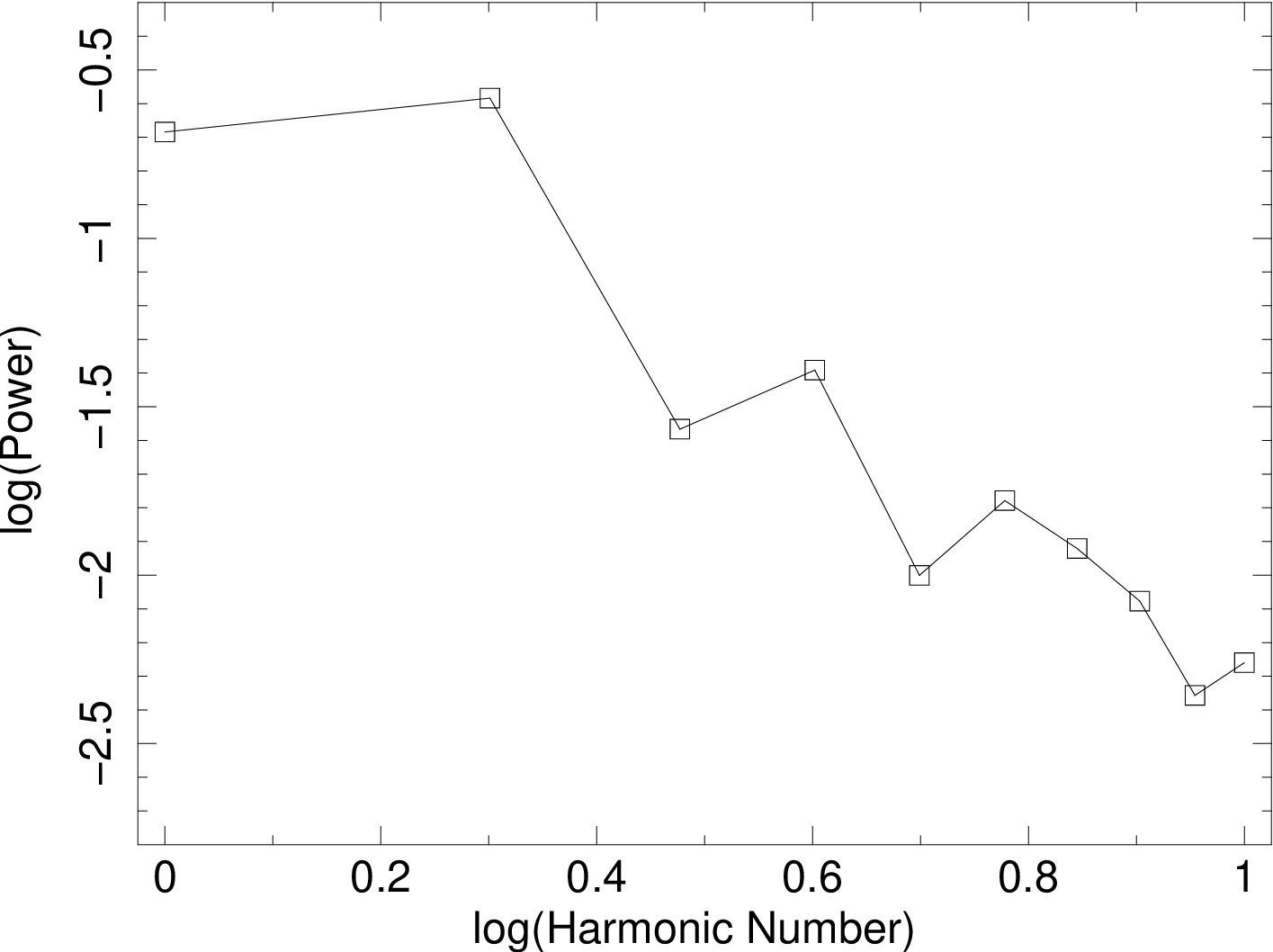} \\
 \end{tabular}}
 \caption{A sample template pulse profile {\bf{(left)}} and its power spectra {\bf{(right)}} in terms of harmonic number obtained from the observation on MJD 55586.}
  \label{sample}
\end{figure}

\begin{figure}
  \center{\includegraphics[width=6.8cm, angle=270]{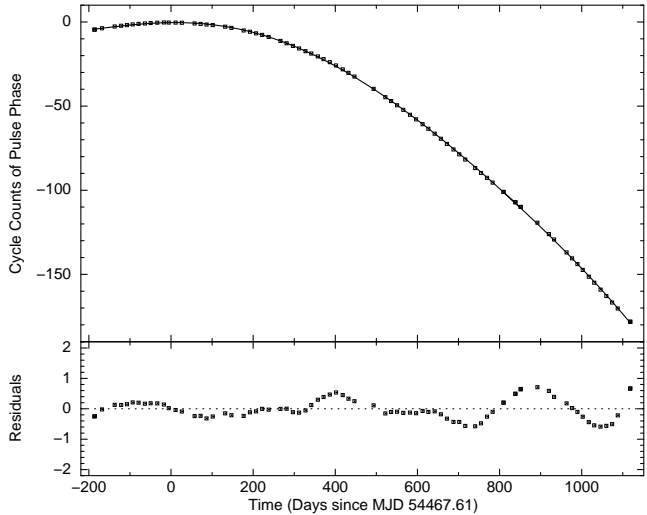}} 
  \caption{Cycle counts of pulse phase and the best-fitting model (solid line). The residuals after the removal of 5th 
  order polynomial (see Equation \ref{polyn}) are given in the bottom panel.}
  \label{cycle}
\end{figure}

\begin{table}
  \caption{Pulse timing solution of 4U 1907+09 between MJD 54280 and 55600.}
  \label{soln}
  \center{\begin{tabular}{lr}
  \hline	
Parameter & Value \\	
 \hline
 Epoch ($t_o$)  			& MJD 54467.61(6) 			  \\
 Pulse Period ($P$)  			& 441.2088(2) s			  	  \\
 Spin frequency ($\nu$)\hspace{1cm}	& $2.2665002(8) \times 10^{-3}$ Hz	  \\
 $\dot{\nu}$ 				& $-3.672(1) \times 10^{-14}$ Hz s$^{-1}$ \\
 $\ddot{\nu}$  				& $-1.497(5) \times 10^{-21}$ Hz s$^{-2}$ \\
 $\dddot{\nu}$  			& $ 1.063(9) \times 10^{-28}$ Hz s$^{-3}$ \\
 $\ddddot{\nu}$  			& $-2.44(3) \times 10^{-36}$  Hz s$^{-4}$ \\
 RMS residual (pulse phase)		& 0.32					  \\
\hline
\end{tabular}} \\
\end{table} 

\subsection{Pulse Frequency Measurements}

For timing analysis, we use 1 s binned background corrected \emph{RXTE}-PCA light curves of the source. These background 
subtracted light curves are corrected to the barycenter of the solar system. Then we correct the light curves for the binary 
motion of 4U 1907+09 using the binary orbital parameters (In't Zand et al. 1998). Since 4U 1907+09 has many dips, we 
eliminate these dips from the light curve.
 Pulse periods for 4U 1907+09 are found by folding the time series on statistically 
independent trial periods (Leahy et al. 1983). Template pulse profiles are constructed from these observations by folding 
the data on the period giving maximum $\chi^2$. The pulse profiles consist of 20 phase bins and are represented by 
their Fourier harmonics (Deeter \& Boynton 1985). We present a sample template pulse profile and its power spectra in
terms of harmonic number in Figure \ref{sample}.

The pulse arrival times are obtained from the cross-correlation between 
template and pulse profiles obtained in each $\sim$2 ks observation. We have been able to connect all pulse arrival times 
of \emph{RXTE} observations in phase over a 1300 day time span.
In the phase connection procedure, in order to
 avoid cycle count ambiguity, we construct the  pulse arrival times
 for a time span where the maximum phase shift is less then 1. This time scale for 4U 1907+09 is around 160 days.
We divide total time span into 10 time intervals intervals each around 160 days and construct
the pulse arrival times with respect to the best period in that time interval. Then,
we align the slopes of pulse arrival times in the overlapping time intervals
and construct pulse arrival times as presented in the upper panel of Figure \ref{cycle}.

 We fit the pulse arrival times to the fifth order polynomial,
\begin{equation}
\delta \phi = \delta \phi_{o} + \delta \nu (t-t_{o})
+ \sum _{n=2}^{5} \frac{1}{n!}
 \frac {d^{n} \phi}{dt^{n}} (t-t_{o})^{n}
\label{polyn}
\end{equation}
where $\delta \phi$ is the pulse phase offset deduced from the pulse timing analysis, $t_{o}$ is the mid-time; 
$\delta \phi_{o}$ is the residual phase offset at t$_{o}$; $\delta \nu$ is the correction to the pulse frequency at time $t_0$; 
$\frac {d^{n} \phi}{dt^{n}} $ for $n$=2,3,4,5 are the first, second, third and fourth order derivatives of the pulse phase. 

The pulse arrival times (pulse cycles) and the residuals of the fit after the removal of the fifth order polynomial trend 
are presented in Figure \ref{cycle}. Table \ref{soln} presents the timing solution of 4U 1907+09 between MJD 54280 and 
55600. It should be noted that we also obtained pulse residuals by performing pulse timing analysis
using the timing solution parameters for pulse frequency and its time
derivatives given in Table \ref{soln}.

\begin{figure*}
  \center{\includegraphics[width=8cm, angle=270]{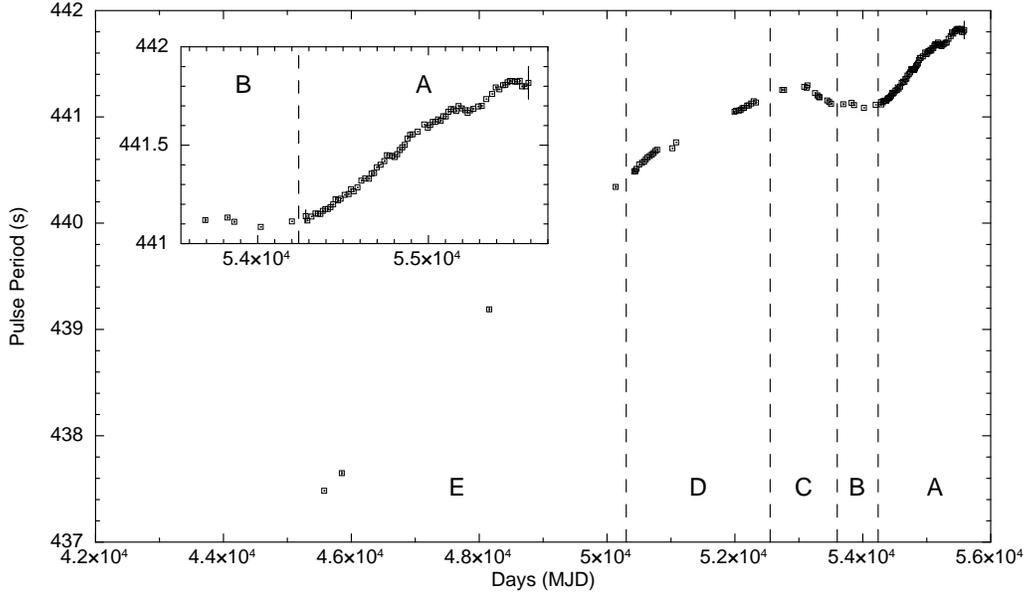}} 
  \caption{Pulse period history of 4U 1907+09. The measurements with \emph{RXTE} and \emph{INTEGRAL} in this work and 
  \.{I}nam et al. (2009a) lie in regions A and B respectively and are separately plotted in the inset (see Table \ref{periods}). Measurements from Fritz et al. (2006) 
  and Baykal et al. (2001, 2006) are in regions C and D respectively. Oldest measurements lie in region E (In't Zand et al. 
  1998, Mihara 1995, Cook \& Page 1987, Makishima et al. 1984).}
  \label{history}
\end{figure*}

In order to obtain pulse frequencies, we fit a linear model to each successive pairs of arrival times. Slopes of these 
linear fits lead us to estimate the pulse frequency values at mid-time of corresponding observations.
In Figure \ref{history} and Table \ref{periods}, we present the pulse period history of the source. In order to 
check the reliability of pulse arrival times, we also obtained the pulse periods using the slopes of pulse
arrivals times obtained from 160-day time intervals. We found exactly the same pulse periods
as obtained from the arrival times of 1300 days time span.
Using the pulse periods, we obtained the pulse frequency value of $\dot \nu = -3.672(1) \times 10^{-14}$ Hz s$^{-1}$ for a time span of $\sim$1300 days .
This is also consistent with the timing solution in Table \ref{soln}.

For the timing analysis of \emph{INTEGRAL} observations, we extract 10 s binned background corrected IBIS-ISGRI light curves 
in the energy range 20 - 40 keV. We sample light curves from $\sim7-10$ days time span and find the best pulse frequency by 
folding the light curve on statistically independent pulse frequencies. Then constructing master and sample pulse profiles, 
we obtain pulse arrival times as described above. From the slopes of pulse arrival times 
($\delta \phi = \delta \nu (t-t_{o})$) we attain the correction for pulse frequencies. We present pulse frequencies 
measured from \emph{RXTE}-PCA and \emph{INTEGRAL} IBIS-ISGRI observations in Figure \ref{history} and Table \ref{periods}. 

\begin{table*}
  \caption{Pulse period measurements of 4U 1907+09.}
  \label{periods}
  \center{\begin{tabular}{cclccl}
  \hline	
Epoch & Pulse Period 	& Reference\hspace{2cm} 	& Epoch & Pulse Period    & Reference	  \\
(MJD) & (s) 		&				& (MJD) & (s)		  &		  \\
 \hline
  53693.0 & 441.1193 $\pm$ 0.0100 & This work$^a$	  &  54877.5 & 441.5344 $\pm$ 0.0035 & This work$^b$ \\
  53823.6 & 441.1322 $\pm$ 0.0070 & This work$^a$	  &  54892.7 & 441.5526 $\pm$ 0.0040 & This work$^b$ \\
  53862.6 & 441.1108 $\pm$ 0.0050 & This work$^a$	  &  54907.0 & 441.5553 $\pm$ 0.0039 & This work$^b$ \\
  54018.1 & 441.0847 $\pm$ 0.0020 & This work$^a$	  &  54937.6 & 441.5687 $\pm$ 0.0012 & This work$^b$ \\
  54200.4 & 441.1127 $\pm$ 0.0014 & This work$^a$	  &  54975.3 & 441.6058 $\pm$ 0.0020 & This work$^b$ \\
  54281.5 & 441.1030 $\pm$ 0.0372 & \.{I}nam et al. 2009a &  54996.5 & 441.5892 $\pm$ 0.0040 & This work$^b$ \\
  54291.0 & 441.1213 $\pm$ 0.0038 & \.{I}nam et al. 2009a &  55010.7 & 441.6029 $\pm$ 0.0039 & This work$^b$ \\
  54315.0 & 441.1367 $\pm$ 0.0021 & \.{I}nam et al. 2009a &  55025.6 & 441.6185 $\pm$ 0.0037 & This work$^b$ \\
  54338.2 & 441.1545 $\pm$ 0.0041 & \.{I}nam et al. 2009a &  55041.4 & 441.6188 $\pm$ 0.0034 & This work$^b$ \\
  54353.3 & 441.1509 $\pm$ 0.0046 & \.{I}nam et al. 2009a &  55057.0 & 441.6310 $\pm$ 0.0038 & This work$^b$ \\
  54367.3 & 441.1543 $\pm$ 0.0047 & \.{I}nam et al. 2009a &  55072.3 & 441.6247 $\pm$ 0.0036 & This work$^b$ \\
  54381.3 & 441.1623 $\pm$ 0.0046 & \.{I}nam et al. 2009a &  55087.2 & 441.6473 $\pm$ 0.0040 & This work$^b$ \\
  54396.2 & 441.1750 $\pm$ 0.0042 & \.{I}nam et al. 2009a &  55102.0 & 441.6462 $\pm$ 0.0037 & This work$^b$ \\
  54410.9 & 441.1761 $\pm$ 0.0047 & \.{I}nam et al. 2009a &  55117.2 & 441.6697 $\pm$ 0.0037 & This work$^b$ \\
  54426.0 & 441.1862 $\pm$ 0.0041 & \.{I}nam et al. 2009a &  55132.0 & 441.6836 $\pm$ 0.0038 & This work$^b$ \\
  54442.1 & 441.1992 $\pm$ 0.0040 & \.{I}nam et al. 2009a &  55147.4 & 441.6826 $\pm$ 0.0035 & This work$^b$ \\
  54456.1 & 441.2245 $\pm$ 0.0056 & \.{I}nam et al. 2009a &  55162.5 & 441.6737 $\pm$ 0.0040 & This work$^b$ \\
  54470.4 & 441.2185 $\pm$ 0.0039 & \.{I}nam et al. 2009a &  55176.8 & 441.6996 $\pm$ 0.0039 & This work$^b$ \\
  54486.3 & 441.2284 $\pm$ 0.0043 & \.{I}nam et al. 2009a &  55196.2 & 441.6874 $\pm$ 0.0023 & This work \\
  54509.4 & 441.2472 $\pm$ 0.0021 & \.{I}nam et al. 2009a &  55215.3 & 441.6777 $\pm$ 0.0040 & This work \\
  54532.3 & 441.2537 $\pm$ 0.0044 & \.{I}nam et al. 2009a &  55229.7 & 441.6647 $\pm$ 0.0039 & This work \\
  54546.8 & 441.2756 $\pm$ 0.0046 & \.{I}nam et al. 2009a &  55243.9 & 441.6772 $\pm$ 0.0041 & This work \\
  54561.6 & 441.2657 $\pm$ 0.0043 & \.{I}nam et al. 2009a &  55264.0 & 441.6848 $\pm$ 0.0021 & This work \\
  54584.3 & 441.2855 $\pm$ 0.0022 & \.{I}nam et al. 2009a &  55291.7 & 441.6973 $\pm$ 0.0020 & This work \\
  54607.1 & 441.3195 $\pm$ 0.0043 & \.{I}nam et al. 2009a &  55312.5 & 441.6998 $\pm$ 0.0043 & This work \\
  54629.6 & 441.3301 $\pm$ 0.0022 & \.{I}nam et al. 2009a &  55339.4 & 441.7343 $\pm$ 0.0014 & This work \\
  54652.1 & 441.3307 $\pm$ 0.0043 & \.{I}nam et al. 2009a &  55373.8 & 441.7607 $\pm$ 0.0020 & This work \\
  54667.2 & 441.3549 $\pm$ 0.0044 & \.{I}nam et al. 2009a &  55394.1 & 441.7943 $\pm$ 0.0045 & This work \\
  54682.1 & 441.3596 $\pm$ 0.0044 & \.{I}nam et al. 2009a &  55415.6 & 441.7842 $\pm$ 0.0018 & This work \\
  54697.5 & 441.3883 $\pm$ 0.0035 & This work$^b$	  &  55437.5 & 441.8052 $\pm$ 0.0042 & This work \\
  54719.7 & 441.4009 $\pm$ 0.0020 & This work$^b$	  &  55450.7 & 441.8073 $\pm$ 0.0043 & This work \\
  54741.6 & 441.4195 $\pm$ 0.0037 & This work$^b$	  &  55463.7 & 441.8199 $\pm$ 0.0043 & This work \\
  54756.8 & 441.4492 $\pm$ 0.0037 & This work$^b$	  &  55477.6 & 441.8271 $\pm$ 0.0038 & This work \\
  54771.9 & 441.4465 $\pm$ 0.0038 & This work$^b$	  &  55491.5 & 441.8263 $\pm$ 0.0043 & This work \\
  54787.2 & 441.4460 $\pm$ 0.0036 & This work$^b$	  &  55505.8 & 441.8235 $\pm$ 0.0037 & This work \\
  54802.1 & 441.4398 $\pm$ 0.0040 & This work$^b$	  &  55520.5 & 441.8227 $\pm$ 0.0040 & This work \\
  54817.2 & 441.4544 $\pm$ 0.0035 & This work$^b$	  &  55534.6 & 441.8278 $\pm$ 0.0040 & This work \\
  54832.1 & 441.4758 $\pm$ 0.0041 & This work$^b$	  &  55548.6 & 441.7993 $\pm$ 0.0041 & This work \\
  54846.6 & 441.4889 $\pm$ 0.0037 & This work$^b$	  &  55570.5 & 441.7994 $\pm$ 0.0019 & This work \\
  54861.8 & 441.5019 $\pm$ 0.0037 & This work$^b$	  &  55585.8 & 441.8172 $\pm$ 0.0834 & This work \\
\hline
\end{tabular}} \\
\begin{flushleft}
$^{a}$ The pulse periods are measured from \emph{INTEGRAL} observations.

$^{b}$ The pulse periods were presented before by \c{S}ahiner et al. (2011).
\end{flushleft}	
\end{table*}

\subsection{Torque Noise Strength}

In order to see the statistical trend of the pulse frequency derivatives, we construct power spectrum of the pulse frequency 
derivatives. We use the Deeter polynomial estimator method (Deeter 1984) to the pulse frequency measurements. This 
technique uses the polynomial estimators instead of sinusoidal estimate for each time scale $T$.

\begin{figure}
  \center{\includegraphics[width=6.2cm, angle=270]{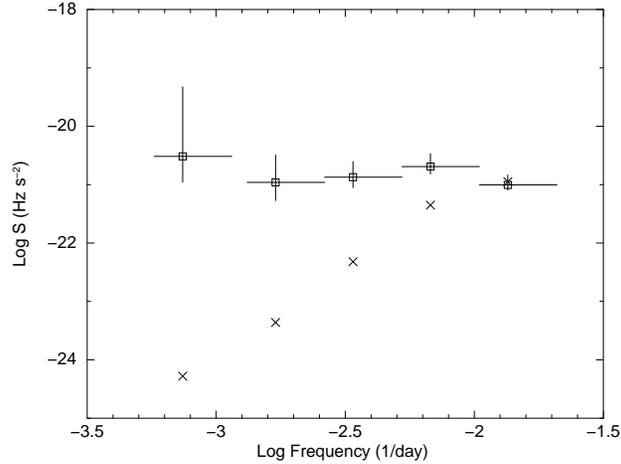}} 
  \caption{Power density of spin frequency derivatives of 4U 1907+09. The power due to measurement noise is subtracted from 
 the estimates and shown independently by the cross symbols. }
  \label{power}
\end{figure}

The power density spectra can be expressed by applying sampling function to the pulse arrival times (Deeter 1984; for the 
applications see Bildsten et al. (1997) and Baykal et al. (2007)). The power density estimator $P_{\dot \nu}({f})$ is 
defined as $\int _{0} ^{\infty} P_{\dot \nu}({f})df = < (\dot \nu - < \dot \nu >)^{2}>$ where $< \dot \nu >$ is the mean 
pulse frequency derivative for a given analysis frequency. In order to estimate the power density, we first divide 
pulse arrival times into time spans of duration $T$ and fit cubic polynomial in time. The observed time series is simulated 
by Monte Carlo techniques for a unit white noise strength defined as $P_{\dot \nu}({f})=1$ and fitted to a cubic polynomial 
in time. Then the square of the third order polynomial term was divided into the value from Monte Carlo simulations 
(Deeter 1984, Cordes 1980). The logarithmic average of these estimators over the same time intervals is the power density 
estimate. This procedure is repeated for different durations $T$ to obtain a power spectrum. We also calculated power density 
with quartic polynomial estimator and found consistent power density spectra with the cubic polynomial estimator. The frequency 
response of each power density and measurement noise estimates are presented at $f \sim 1/T$. In Figure \ref{power}, we 
present power of pulse frequency derivatives ($P_{\dot \nu}({f})$) per Hertz as a function of analysis frequency $f$. The 
slope of the power spectrum between 1/1300 and 1/75 d$^{-1}$ is flat. This suggests that fluctuations in the pulse frequency 
derivatives are white noise and the pulse frequency fluctuations are consistent with random walk model. The noise strength 
is found to be $1.27\times10^{-21}$ Hz s$^{-2}$.

 Random walk model in pulse frequency or white noise model in  
pulse frequency derivative are appropiate models for wind accretors e.g. Vela X-1, 4U 1538-52 and GX 301-2 (Deeter 1981, 
Deeter et al. 1989, Bildsten et al. 1997). They have flat power spectra with white noise strength in the range 
$10^{-20}-10^{-18}$ Hz s$^{-2}$. However, Her X-1 and 4U 1626-67 are disc accretors with low mass companions which have shown 
pulse frequency time series consistent with the random walk model (Bildsten et al. 1997). Their white noise strengths are in the 
range $10^{-21}$ to $10^{-18}$ Hz s$^{-2}$. In these systems red noise in pulse frequency can not be ruled out (Bildsten et 
al. 1997). For disc accretors, Cen X-3 have red noise in pulse frequency derivatives, noise strength varies from low to high 
frequencies as $10^{-16}$, $10^{-18}$ Hz s$^{-2}$ (Bildsten et al. 1997). Power law index of power spectra in this 
system is $\sim-1$. This implies that, at short time scales, disc accretion is dominated and noise is less, on the other hand at 
time scales longer than viscous time scales there are excessive noise.

\section{Spectral Analysis}

\subsection{Orbital Phase Resolved Spectra with \emph{RXTE}-PCA}{\label{specsect}}

The \emph{RXTE} observations of 4U 1907+09 were proposed with a time distribution such that they spread over a variety of 
orbital phases, therefore they are very suitable for the investigation of the variation in the spectral parameters through 
the binary orbit. During the analysis of PCA spectra energy range is restricted to 3 - 25 keV since the count statistics 
is poor beyond this range. The spectral analysis is carried out by using the \verb"XSPEC v.12.6.0" software. A systematic 
error of 2\% is applied to handle the uncertainties in the response matrices and in background modelling (Wilms et al. 1999).

4U 1907+09 is near the Galactic plane and the supernova remnant W49B, therefore an extra process is needed for a correct 
estimation of the background spectra. The exceptional dipping states of the source provide a good basis for background 
estimation since the count rates are consistent with the Galactic ridge emission (Roberts et al. 2001, Baykal et al. 2006). 
4U 1907+09 is detected to be in the dipping state in about 55 ks of the 216 ks \emph{RXTE}-PCA observations. The overall 
dipping state spectrum is used as an additional background, the procedure was explained before by \.{I}nam et al. (2009a).

\begin{figure}
  \center{\includegraphics[width=9.1cm, angle=270]{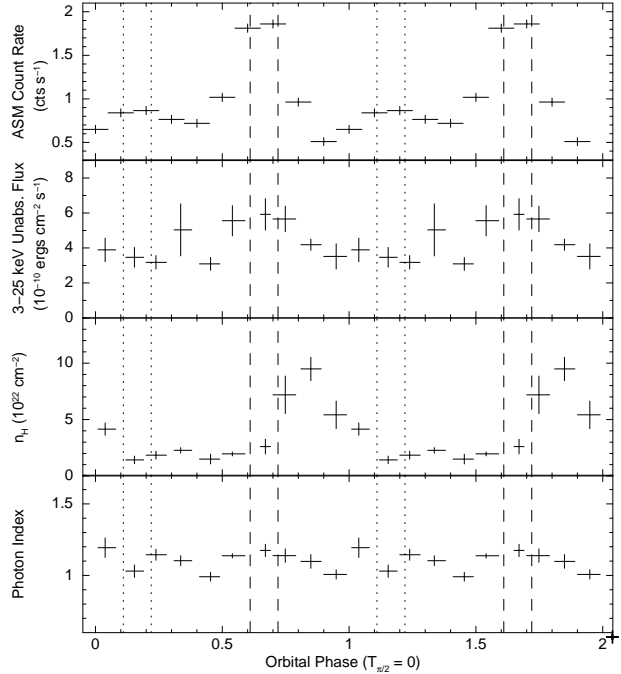}} 
  \caption{Variations of spectral parameters through binary orbit. The data points are repeated for a cycle for clarity. 
From top to bottom; \emph{RXTE}-ASM folded orbital profile, unabsorbed flux at 3 - 25 keV, Hydrogen column density and 
  photon index plotted over orbital phase presented with 10 bins respectively. All uncertainties are calculated at 90\% confidence level. The 
  vertical dashed lines in all panels correspond to time of periastron passage vertical dotted lines in all panels correspond to time of apastron passage  within 1 $\sigma$.}
  \label{spe}
\end{figure}

A total of 71 non-dip spectra of 4U 1907+09 are analysed for the orbital variation of the spectral parameters. Results of 
first 19 spectra were also given in the previous paper (\.{I}nam et al. 2009a). The basic model which consists of a power law (\verb"powerlaw")
with a high energy cutoff (\verb"highecut") and photoelectric absorption (\verb"wabs") is successful for only 10 spectra, where as an additional model  
component for the CRSF around 19 keV is required for the rest. We use \verb"cyclabs" model in \verb"XSPEC" 
(Mihara et al. 1990 and Makishima et al. 1990). The fundamental line energy is fixed at 18.9 keV  due to the statistical insignificance of individual PCA spectra with short exposure time ( $\sim$2 ks) and the mean values of the 
depth and the width of the line are found as 0.4 keV and 1.7 keV respectively. These values are used  to model CRSF since they agree within 1 
$\sigma$ uncertainty with the previous CRSF parameters (Mihara 1995, Cusumano et al. 1998 and Makishima et al. 1999). Addition of a model component for the weak \emph{Fe} emission line at 6.4 keV did not improve our fits, 
since this feature had lost its strength after subtraction of the diffuse emission from the Galactic ridge.

Orbital variations of unabsorbed flux at 3 - 25 keV, Hydrogen column density (n$_H$) and photon index are plotted in Figure 
\ref{spe}, which is an updated version of Figure 5 in \.{I}nam et al. (2009a). The variability of n$_H$ over the binary orbit 
is evident in the third panel of Figure \ref{spe}. The base value of n$_H$ is $\sim2 \times 10^{22}$ cm$^{-2}$. It increases 
up to a value of $\sim10.5 \times 10^{22}$ cm$^{-2}$ just after the periastron passage (indicated by the vertical dash lines)
and it remains at high values until the apastron, where it reduces to its base value again. 9 of our individual n$_H$ measurements 
exceed previously reported maximal value ($\sim9 \times 10^{22}$ cm$^{-2}$, In't Zand et al. 1997). Similar orbital dependence of $n_H$
was also reported for the source before (Roberts et al. 2001). One can see that there is 
no significant orbital variation of photon index in the bottom panel of Figure \ref{spe}, the mean value is $\sim$1.1. Other 
parameters that are consistent with being constant through the orbit are high energy cutoff and exponential folding energy 
with the mean values 12.7 keV and 9.6 keV respectively. 

In some accretion powered pulsars (e.g. Vela X-1 (Haberl \& White 1990)) n$_H$ is highly variable over the orbital phases and 
can range up to $\sim$10$^{23}$ - 10$^{24}$ cm$^{-2}$. This variation is a probe to the density distribution of the 
accreting matter. The observed increase in absorption at periastron passage or eclipse ingress may be due to a gas stream 
from the companion trailing behind the pulsar (Haberl et al. 1989). In the case of 4U 1907+09, n$_H$ increasing to its 
maximum value after the periastron passage implies that the location of the absorbing material is the dense stellar wind of 
the companion star (Roberts et al 2001). Leahy (2001) and Kostka \& Leahy (2010) modeled the absorption profile according to 
theoretical wind models and they have proposed that the most probable mechanism for 4U 1907+09 is accretion from a spherical 
wind with equatorially enhanced dense spiral stream of gas around the companion star.

\subsection{X-ray Flux Dependence of Spectral Parameters}

\begin{figure}
  \center{\includegraphics[width=6.2cm, angle=270]{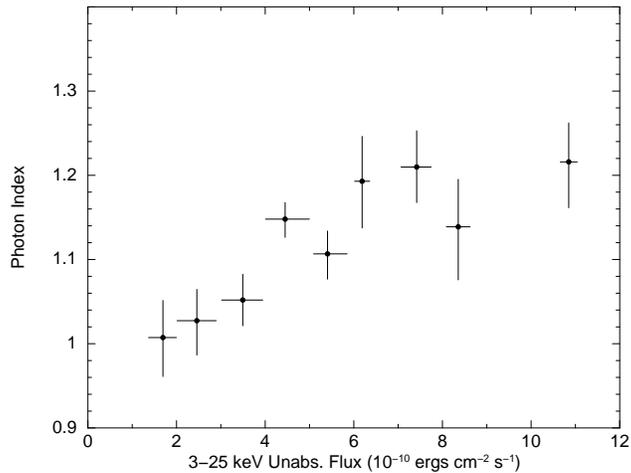}} 
  \caption{Variation of power law index with X-ray flux from \emph{RXTE-PCA} observations.}
  \label{flux_index}
\end{figure}

To look for the X-ray flux dependence of spectral parameters of the source, we use spectral parameters and X-ray flux values obtained from 71 spectra that are presented in Section \ref{specsect}. We sorted observations according to the X-ray flux values and spectral parameters of the observations with similar X-ray flux values are averaged to obtain a spectral parameter set as a function of X-ray flux. It should be noted that range of each flux value over which the spectral parameters are averaged is represented as x-axis error bars of the data points in Figure \ref{flux_index}. We find that no spectral parameters except power law index show any correlation with the 3-25 keV unabsorbed X-ray flux. Whereas power law index is found to be correlated with the X-ray 
flux (see Figure \ref{flux_index}). 

Correlation between power law index and X-ray flux is an indication of spectral hardening with decreasing X-ray flux. Similar 
correlation has been seen in the accretion powered pulsars V0332+53 (Klochkov et al. 2011), 4U 0115+63 (Klochkov et al. 2011), 
4U 1626-67 (Jain et al. 2010) and Vela X-1 (F\"{u}rst et al. 2009). On the other hand, there are also some accretion powered 
pulsars that exhibit spectral softening with decreasing X-ray flux: Her X-1 (Kolochkov et al. 2011), Swift J1626.6-5156 
(\.{I}\c{c}dem et al. 2011), XMMU J054134.7-682550 (\.{I}nam et al. 2009b), 2S 1417-62 (\.{I}nam et al. 2004) and 
SAX J2103.5+4545 (Baykal et al. 2007). Both correlation and anti-correlation of power law index with X-ray flux might be 
considered as results of mass accretion rate variations and/or inhomogeneities in the companion's wind.

\section{Dipping States and Pulse-to-pulse Variability}{\label{p2p}}

Since 4U 1907+09 is known to be a variable X-ray source, we analysed light curve of each \emph{RXTE}-PCA observation to 
investigate pulse-to-pulse variations.The source is known to have dipping episodes that show no pulsations (In't Zand et al. 
1997); therefore we searched for irregularities of pulsations due to dipping ingress and egress. Although short exposure time 
($\sim$2 ks) of \emph{RXTE} observations allow us to see at most about five successive pulses, we were able to detect a 
variety of rapid pulse shape changes. 

\begin{figure}
  \center{\includegraphics[width=20.4cm, angle=270]{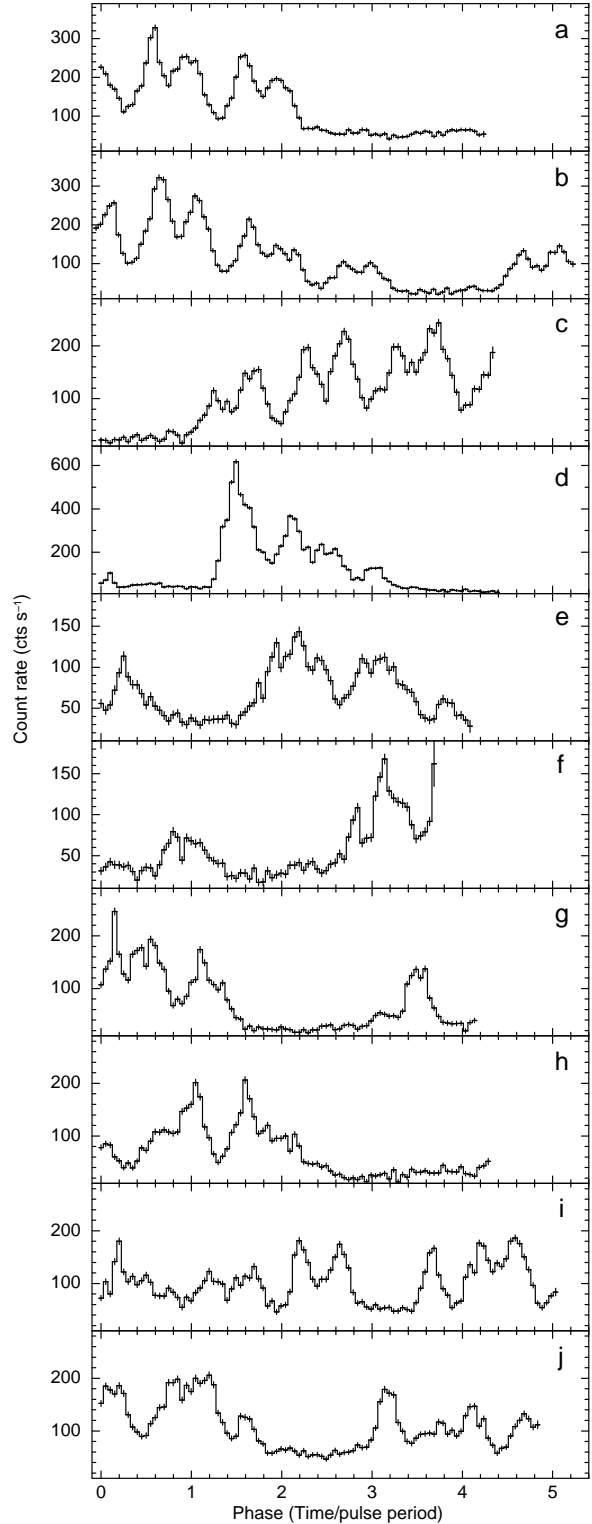}} 
  \caption{Sample set of 22 s binned light curves of different \emph{RXTE}-PCA observations display the variations of pulse 
  shape from pulse-to-pulse and the irregularities due to dipping activity. Time values are converted to phases (or 
  time/pulse period) for arbitarly observation epochs. Explanations to each panel are given in Section \ref{p2p}. Observation 
  IDs of panels from top to bottom are a:94036-01-19-00, b:96366-02-01-00, c:94036-01-05-00, d:94036-01-23-00, 
  e:94036-01-13-00, f:93036-01-20-00, g:94036-01-03-00, h:93036-01-34-00, i:93036-01-38-00 and j:95350-01-06-00.}
  \label{varpulse}
\end{figure}

In Figure \ref{varpulse}, an example of a pulsating and dipping light curve of 4U 1907+09 is given in panel (a). Generally, 
the pulse-to-pulse X-ray intensity decreases just before dipping ingress (see panel (b)) and it increases gradually after the 
dipping egress (see panel (c)). However this is not the case for every dipping episode. The light curve in panel (d) is an 
extreme example, in which an X-ray flare follows a dipping episode and after a smooth decrease for two pulses, the source 
undergoes another dip again. Although the pulse shape is conserved in these primary examples, we find that it can also be 
totally disrupted by the dip, e.g. in panel (e) the pulse shape transforms to a single broad peak after a one pulse long dip 
and in panel (f) the attached wide peaks of the faint pulse are converted into detached sharp peaks after the dip. The 
beginning of the light curve in panel (f) is also an example of unsuccessful dip which may be observed occasionally (see also 
In't Zand et al. 1997). In panel (g), single peak pulses are observable before and after the 1.5 phase long dip. In panel (h), 
after two broad pulses a dipping episode starts on phase 2.6 and continues until the end of the observation. A very rare 
condition is the total disappearance of only one peak of the pulse in panel (i), this example is the shortest dip we observe 
(between phases 2.9 and 3.4). We observe one pulse dips more commonly e.g. in panel (j) the dip between phases 1.8 and 2.8.

\begin{figure}
  \center{\includegraphics[width=6.1cm, angle=270]{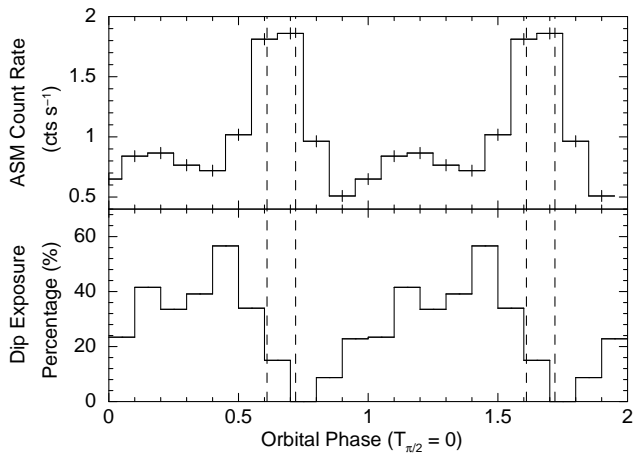}} 
  \caption{ASM light curve folded at orbital period (upper panel) and percentage of dip exposure times to the total exposure 
  through the binary orbit (lower panel). The data points are repeated for a cycle for clarity. The vertical dashed lines 
  correspond to time of periastron passage within 1 $\sigma$. It is evident that probability of observing a dipping state is 
  lower during periastron passage and afterwards (between orbital phases 0.6 and 0.9).}
  \label{dipper}
\end{figure}

Dipping states observed in Vela X-1 (Kreykenbohm et al. 2008) have no identifiable transition phase in spectra before or 
after the dip. However, a recent study of GX 301-2 (G\"{o}\u{g}\"{u}\c{s} et al. 2011) reported a spectral softening starting 
before the dip. The photon index values reaching maximum during the peculiar dip of GX 301-2 resumes its normal values 
immediately after the dip. F\"{u}rst et al. (2011) studied a dip spectra of GX 301-2 and found that although the pulses cease almost completely, $n_H$ does not show any significant variation.  In the case of 4U 1907+09, direct spectral study of the dips with \emph{RXTE} is not possible due to the 
diffuse galactic emission background, since the source is below the detection threshold and the dip spectra are similar to 
the Galactic Ridge spectra. However, it appears from Figure \ref{dipper} that the source may be observed in dipping state in every orbital phase except between the phases 0.7 - 0.8 . Comparing this result with the spectral results of non-dipping observations (see Figure \ref{spe}); 
the absence of dips matches with the orbital phases when n$_H$ reaches to its maximum value just after the periastron passage 
(see lower panel of Figure \ref{dipper}). Moreover, the possibility of dipping is less in orbital phases from 0.8 to 1.1 when 
compared to phases from 0.1 to 0.6; which is in anti-correlation with the n$_H$ values. The source frequently undergoes in 
dipping state during the times with minimum n$_H$ and the possibility peaks between the phases 0.4 - 0.5. It is important to note that this is not a strong conclusion,  since direct measurement of the spectral parameters of the dips is not possible for us.    

Recently, strongly structured wind of the optical companion is accused of being the reason of the 
dips in HMXB light curves. The clumpy wind of the companion, being highly inhomogeneous, creates regions of different density. 
Recent models have shown that the density of the wind material can change up to several orders of magnitude (Runacres \& 
Owocki 2005). Therefore, a sudden decrease of X-ray luminosity is caused by the decrease in mass accretion rate. As the 
accretion rate decreases the Alfv\'{e}n radius increases and accretion stops when the source enters to propeller regime. 
Hence, the disappearance of pulsations can be explained by the passage of the pulsar through low density regions of the wind. 
The limit to the luminosity that turn on the propeller mechanism can be derived from the condition where the magnetospheric 
radius is equal to the co-rotation radius (Illarionov \& Sunyaev 1975) and it is $6 \times 10^{31}$ erg s$^{-1}$ for the case of 
4U 1907+09. Therefore, the dipping states are candidate episodes for propeller regime since the count rates are below 
detection threshold. Similar propeller transition is suggested as the cause of observed dips in the accretion powered pulsars 
GX 1+4 and GRO J1744-28 (Cui 1997).

The dipping episodes of 4U 1907+09 are known to have irregular time coverage, with varying duration. We investigate the 
distribution of dipping states through the orbital phase. About 46 of the 98 observations include dipping states. The 
percentage of exposure time spend in dipping is $\sim$28\% of the total exposure. In order to obtain the orbital distribution 
of dipping states, we calculate the percentage of the dip exposure to total exposure corresponding to each orbital phase with 
0.1 phase resolution. The variation of the dip exposure percentage on orbital phase is plotted together with folded ASM light 
curve at orbital period at Figure \ref{dipper}. It is clearly seen that the occurrence of dips sharply decreases during the 
periastron passage and after the passage. Around the periastron passage, ASM count rate and therefore X-ray flux increases, whereas number of 
dip states decreases. It should be kept in our mind that a transient QPO had been 
found during one of the flares which suggests the formation of a transient disc (Int Zand et al. 1997). This implies that the 
number of occurrences of dip states is less at the periastron passage and afterwards since the disc accretion is dominant 
mechanism. After the apastron (phases from 0.1 to 0.5), number of dip states increases due to the clumpy nature of accretion and occurrence of 
transient propeller regime episodes increases. The latter naturally explains the variable nature of the spin down. The 
combination of disc and wind accretion at different phases of orbit may lead to random walk nature of pulse frequency time 
series.   

\section{Summary}

In this paper, we present analysis of \emph{INTEGRAL} IBIS-ISGRI (between 2005 October and 2007 November) and \emph{RXTE}-PCA 
(between 2007 June and 2011 March) observations of the accretion powered pulsar 4U 1907+09. 

Using \emph{INTEGRAL} and \emph{RXTE} observations, we report new pulse period measurements and obtain an updated version 
of the pulse period history of \.{I}nam et al. (2009a) which is shown in Figure \ref{history}. Using these measurements, we 
construct power spectrum density of the pulse frequencies (see Figure \ref{power}). We find that fluctuations in the pulse 
frequency derivatives are white noise and the pulse frequency fluctuations are consistent with the random walk model. The noise 
strength is found to be $1.27\times 10^{-21}$ Hz s$^{-2}$. We infer that in short time scales, there may be transient disc 
formation around the neutron star which causes random walk in the pulse frequency while in the long term the spin down rate is 
steady. 

From the X-ray spectral analysis of \emph{RXTE}-PCA observations, we find that only Hydrogen column density (n$_H$) is 
significantly variable over the binary orbit, tending to increase just after the periastron passage (see Figure \ref{spe}). 
This might indicate that the location of the absorbing material is the dense stellar wind of the companion star. We also 
study flux dependence of spectral parameters and find that power law index show correlation with the 3-25 keV unabsorbed 
X-ray flux indicating spectral hardening with decreasing X-ray flux (see Figure \ref{flux_index}). This spectral variation 
might be related to mass accretion rate variations and/or inhomogeneities in the companion's wind.

We also look for the irregularities of pulsations near dipping ingress and egress (see Figure \ref{varpulse}). We propose that 
the disappearance of pulsations can be explained by the passage of the pulsar through low density regions of the clumpy wind 
which may lead to a turn-on of a temporary propeller state so that the X-ray luminosity decreases below the detection threshold. 
We are able to analyse orbital dependence of the dip state occurrences (see Figure \ref{dipper}). We find that the source more 
likely undergoes in dipping state after the apastron until the periastron where transitions to temporary propeller state might occur due 
to accretion from clumpy wind. 

\section*{Acknowledgment}

We acknowledge support from T\"{U}B\.{I}TAK, the Scientific and Technological Research Council of Turkey through the 
research project TBAG 109T748. We thank Prof. \"{U}mit  K{\i}z{\i}lo\u{g}lu and Mehtap \"{O}zbey for useful discussions.

\bsp

\label{lastpage}


\begin{thebibliography}{99}

\bibitem{}
Baykal A., \.{I}nam S. \c{C}., Alpar M. A., In't Zand J., Strohmayer T., 2001, MNRAS, 327, 1269
\bibitem{}
Baykal A., \.{I}nam S. \c{C}., Beklen E., 2006, MNRAS, 369, 1760
\bibitem{}
Baykal A., \.{I}nam S. \c{C}., Stark M. J., Heffner C. M., Erkoca A. E., Swank J. H., 2007, MNRAS, 374, 1108
\bibitem{}
Bildsten L., Chakrabarty D., Chiu J. et al., 1997, ApJS, 113, 367
\bibitem{}
Chitnis V. R., Rao A. R., Agrawal P. C., Manchanda R. K., 1993, A\&A, 268, 609
\bibitem{}
Coburn W., Heindl W. A., Rothschild R. E. et al., 2002, ApJ, 580, 394
\bibitem{}
Cook M. C., Page C. G., 1987, MNRAS, 225, 381
\bibitem{}
Corbet R. H. D., 1984, A\&A, 141, 91
\bibitem{}
Cordes J. M., 1980, ApJ, 237, 216
\bibitem{}
Cox N. L. J., Kaper L., Mokiem M. R., 2005, A\&A, 436, 661
\bibitem{}
Cui W., 1997, ApJ, 482, L163
\bibitem{}
Cusumano G., di Salvo T., Burderi L., Orlandini M., Piraino S., Robba N., Santangelo A., 1998, A\&A, 338, L79
\bibitem{}
Deeter J. E., 1981, Ph.D. Thesis, Washington Univ., Seattle
\bibitem{}
Deeter J. E., 1984, ApJ, 281, 482
\bibitem{}
Deeter J. E., Boynton P. E., 1985, in Hayakawa S. and Nagase F., Proc. Inuyama Workshop: Timing Studies of X-Ray Sources, 
p.29, Nagoya Univ., Nagoya
\bibitem{}
Deeter J. E., Boynton P. E., Lamb F. K., Zylstra G., 1989, ApJ, 336, 376
\bibitem{}
Fritz S., Kreykenbohm I., Wilms J. et al., 2006, A\&A, 458, 885
\bibitem{}
F\"{u}rst F., Kreykenbohm I., Wilms J., Kretschmar P., Klochkov D., Santangelo A., Staubert R., 2008, Conf. Proc. of 7th 
INTEGRAL workshop - An INTEGRAL View of Compact Objects Proceedings, PoS Integral08:119
\bibitem{}
F\"{u}rst F. et al. 2011, A\&A, 535, 9
\bibitem{}
Giacconi R., Kellogg E., Gorenstein P., Gursky H., Tananbaum H., 1971, ApJ, 165, L27
\bibitem{}
G\"{o}\u{g}\"{u}\c{s} E., Kreykenbohm I., Belloni T. M., 2011, A\&A, 525L, 6
\bibitem{}
Haberl F., White N. E., 1990, ApJ, 361, 225
\bibitem{}
Haberl F., White N. E., Kallman T. R., 1989, ApJ, 343, 409
\bibitem{}
Illarionov A. F., Sunyaev R. A., 1975, A\&A, 39, 185
\bibitem{}
\.{I}\c{c}dem B., \.{I}nam S. \c{C}., Baykal A. 2011,  MNRAS, 415, 1523 
\bibitem{}
\.{I}nam S. \c{C}., Baykal A., Scott D.M., Finger M., Swank J., 2004, MNRAS, 349, 173
\bibitem{}
\.{I}nam S. \c{C}., \c{S}ahiner \c{S}., Baykal A., 2009a, MNRAS, 395, 1015
\bibitem{} 
\.{I}nam S. \c{C}., Townsend L. J., McBride V. A., Baykal A., Coe M. J., Corbet R. H. D., 2009b, MNRAS, 395, 1662
\bibitem{}
In't Zand J. J. M., Strohmayer T. E., Baykal A., 1997, ApJ, 479, L47
\bibitem{}
In't Zand J. J. M., Baykal A., Strohmayer T. E., 1998, ApJ, 496, 386
\bibitem{}
Iye M., 1986, PASJ, 38, 463
\bibitem{}
Jahoda K., Swank J. H., Giles A. B., Stark M. J., Strohmayer T., Zhang W., Morgan E. H., 1996, Proc. SPIE, 2808, 59
\bibitem{}
Jain C., Paul B., Dutta, A., 2010, MNRAS, 403, 920
\bibitem{}
Klochkov D., Santangelo A., Staubert R., Rothschild R. E., 2011, Conf. Proc. of 8th INTEGRAL Workshop, The Restless Gamma-ray 
Universe, 2010, Dublin, Ireland; arXiv:1105.3547
\bibitem{}
Kostka M., Leahy D. A., 2010, MNRAS, 407, 1182
\bibitem{}
Kreykenbohm I., Wilms J., Kretschmar P. et al., 2008, A\&A, 492, 511
\bibitem{}
Leahy D. A., Darbro W., Elsner R. F., Weisskopf M. C., Kahn S., Sutherland P. G., Grindlay J. E., 1983, ApJ, 266, 160
\bibitem{}
Leahy D. A., 2001, Proc. 27th International Cosmic Ray Conference, p.2528, Copernicus Gesellschaft, Hamburg
\bibitem{}
Lebrun F., Leray J. P., Lavocat P. et al., 2003, A\&A, 411, L141
\bibitem{}
Makishima K., Kawai N., Koyama K., Shibazaki N., 1984, PASJ, 36, 679
\bibitem{}
Makishima K., Ohashi T., Kawai N. et al., 1990, PASJ, 42, 295
\bibitem{}
Makishima K., Mihara T., Nagase F., Tanaka Y., 1999, ApJ, 525, 978
\bibitem{}
Marshall N., Ricketts M. J., 1980, MNRAS, 193, P7
\bibitem{}
Mihara T., 1995, PhD thesis, RIKEN, Tokyo
\bibitem{}
Mihara T., Makishima K., Ohashi T., Sakao T., Tashiro M., 1990, Nature, 346, 250
\bibitem{}
Mukerjee K., Agrawal P. C., Paul B., Rao A. R., Yadav J. S., Seetha S., Kasturirangan K., 2001, ApJ, 548, 368
\bibitem{}
Nespoli E., Fabregat J., Mennickent R. E., 2008, A\&A, 486, 911
\bibitem{}
Rivers E., Markowitz A., Pottschmidt K. et al., 2010, ApJ, 709, 179
\bibitem{}
Roberts M. S. E., Michelson P. F., Leahy D. A., Hall T. A., Finley J. P., Cominsky L. R., Srinivasan R., 2001, ApJ, 555, 967
\bibitem{}
Runacres M. C., Owocki S. P., 2005, A\&A, 429, 323
\bibitem{}
\c{S}ahiner \c{S}., \.{I}nam S. \c{C}., Baykal, A., 2011, AIPC, 1379, 214
\bibitem{}
Schwartz D. A., Griffiths R. E., Bowyer S., Thorstensen J. R., Charles P.A., 1980, AJ, 85, 549
\bibitem{}
Ubertini P., Lebrun F., Di Cocco G. et al., 2003, A\&A, 411, L131
\bibitem{}
Wilms J., Nowak M. A., Dove J. B., Fender R. P., Di Matteo T., 1999, ApJ, 522, 460 

\end{thebibliography}
\end{document}